\documentstyle[sprocl]{article}

\newcommand{\etal}{\hbox{et~al.}}

\begin{document}

\title{A NEW VISION OF THE COMA CLUSTER: \\ CONFERENCE SUMMARY}

\author{MATTHEW COLLESS}

\address{Mount Stromlo and Siding Spring Observatories\\
         The Australian National University\\
         Weston Creek, Canberra, ACT 2611, Australia\\
         E-mail: colless@mso.anu.edu.au}

\maketitle
\vspace*{1cm}

\section{Introduction}

In his introductory historical review, {\sc Biviano} provided a very
thorough precis of previous studies of the Coma cluster stretching
back to the beginning of this century and forward to the last couple
of years (gracefully allowing the protagonists present at the meeting
to present their own versions of the most recent work). He quantified
Coma's well-established position in the astronomical literature as the
archetypal rich cluster, finding that papers on Coma represent about
5\% of all papers on clusters of galaxies. He also emphasised that the
last decade has seen an acceleration in the number of papers on Coma
published per year. This renewed interest in the Coma cluster is
reflected in the number of people attending this conference and in the
extraordinary breadth of the research with Coma as its focus that was
reported here. The current vitality of cluster studies seems to be
based in part on technological advances which have produced high
quality data at wavelengths outside the visible and led to a far
fuller and richer picture of galaxy clusters, and in part on renewed
interest in clusters as structured, forming entities and as
laboratories for understanding galaxy formation and interaction
processes in extreme density environments.

Given this history, it seems surprising that there has never
previously been any conference focussed solely on Coma. This meeting
has rectified that deficiency, and provided an inclusive, detailed and
up-to-date snapshot of research on Coma which will serve as an
invaluable source-book and inspiration for future work. In this
summary I attempt to draw together a few of the highlights and major
themes of the conference.

\section{Substructure and Superstructure}

The review by {\sc West} of the Coma cluster in relation to its
large-scale environs offered a view from the top of a bottom-up
universe. He stressed the observational evidence for strong cross-talk
between structures over a range of 10$^4$ in size, from individual
galaxies on scales of 10\,kpc, through clusters with sizes of order
1\,Mpc, up to `Great Filaments' extending as much as 100\,Mpc. His theme
was that ``filaments drive cluster formation'', and he was able to
show impressive evidence for this view, including the remarkable
alignment of structures on all scales: the optical isophotes of the
dominant galaxies, the overall distributions of cluster galaxies and
X-ray gas, and the great chain of galaxies extending across the sky,
passing through both Coma and Abell 1367, which is usually called the
Great Wall (though he argued its axial ratios make that a misnomer,
and it should be called the Great Filament). This evidence, {\sc West}
contended, meant that Coma was built by the infall of subclusters
along the Great Filament.

This embedding of Coma within the surrounding superstructure has been
complemented by a growing understanding of the internal substructure
of the cluster. Up to the early 1980's dynamical studies of Coma
tended to emphasise the relaxed, regular nature of the cluster in
order to make possible analytical attacks on the hallowed questions
regarding the amount and distribution of the cluster's dark matter,
famously discovered by Zwicky (1933). The late 1980's saw a change of
view, as various pieces of evidence showed that Coma was clearly not
spherically symmetric and dynamically fully relaxed. The last few
years have seen further rapid progress in this direction, most notably
in the deep ROSAT X-ray maps of the cluster (White \etal\ 1993) and
dynamical analyses based on many hundreds of galaxy velocities
(Colless \& Dunn 1996; Biviano \etal\ 1996).

With cluster structure now confirmed and quantified, the emphasis is
switching to understanding the irregular dynamics and history of
Coma's formation. This was reflected both in {\sc West}'s review and
in the results presented by {\sc van Haarlem}, who is measuring
velocities for the galaxies in the outer parts of the cluster to
obtain a fuller picture of the anisotropic infall onto the
cluster. Other relevant new results in this area were presented by
{\sc Gerbal}, {\sc Slezak} and {\sc Gurzadyan}. {\sc Gerbal} and {\sc
Slezak} both presented wavelet analyses of cluster substructure and
showed that the bright galaxies in the cluster core form two tight
groups around the the two dominant galaxies, NGC\,4874 and NGC\,4889;
however the fainter galaxies are centred in-between, and are more
smoothly distributed over the cluster. {\sc Gerbal} and {\sc
Gurzadyan} presented different methods for decomposing the cluster
into a hierarchy of dynamically significant sub-units, the former
using a binding energy criterion, the latter using Riemann curvature
to quantify the boundness of orbits in phase space. These
investigations confirmed that the group of galaxies around NGC\,4889 is
undergoing disruption as suggested by Colless \& Dunn (1996), but, in
a modification of that picture, envisage the main cluster being traced
by the fainter galaxies, with NGC\,4874 and its associated bright
galaxies another embedded subcluster. On the question of whether the
dominant galaxies are a bound pair, however, the two analyses
disagreed.

One issue that was repeatedly raised during the conference was whether
the subcluster around the cD galaxy NGC\,4839, lying 40\,arcmin away
from the cluster centre towards the south-west, has already passed
through the cluster core or not. Burns \etal\ (1994) argued that the
distribution of X-ray gas and galaxies was more consistent with a
group that had been partially disrupted on a first passage through the
cluster, whereas Colless \& Dunn (1996) marshalled a variety of
evidences pointing towards the group being on its first approach to
the cluster. In fact the question of whether the NGC\,4839 group has or
has not already passed through the cluster core is, in itself, only of
limited interest. However, that so much attention should be directed
towards this question does illustrate the extent to which the primary
goals for dynamical analysis of clusters have broadened from the
issues of the distribution of luminous and dark matter to encompass
the broader questions regarding the formation history of the cluster.

Some radio observations highly pertinent to this question were
summarised at this meeting by {\sc Feretti}. She showed that NGC\,4839
is a wide-angled tail (WAT) radio source with the tail trailing away
from the cluster centre. Since NGC\,4839 appears to be at rest within
its group, this implies that the WAT is produced as the group passes
through the intra-cluster medium of the main cluster and hence that it
is either on first approach or falling back in after having passed
through the main cluster. The radio halo of the Coma cluster (an
attribute seen in only 10 clusters) has a lifetime of around 0.1\,Gyr,
whereas the NGC\,4839 group is about 1\,Gyr travel-time from the cluster
core. A more plausible energy source for the radio halo is thus the
ongoing merger between the NGC\,4874 and NGC\,4889 subclusters. This is
supported by the observation that the halo's spectral index is
$\alpha$$\approx$0.8 in the centre and higher outside, implying the
energy source is within the optical core radius of the cluster.

Some puzzles regarding the radio observations of Coma remain,
however. The mechanism by which the kinetic energy of a subcluster
merger is converted into the luminosity of the radio halo remains
unclear; since such halos are rare yet subcluster mergers seem
relatively common, perhaps special conditions are required. The origin
of the apparent bridge between the halo and NGC\,4839 seen in the radio
map of Deiss \etal\ (1997) is not known; it cannot be due to the
passage of that subcluster since again the timescale is wrong. Finally
there is the intriguing extended radio source 1253+275, which lies
along the NGC\,4874/NGC\,4839 axis but half as far again from the
cluster centre.

\section{The Properties of Galaxies in Coma}

Even more strongly than the dynamical analyses, the work on the
properties of Coma cluster galaxies showed the new impetus provided by
multi-waveband studies. This was exemplified in the review by {\sc
Gavazzi}, who explored a wide range of galaxy properties, comparing
cluster galaxies to field galaxies, using radio, HI, CO, far-infrared,
near-infrared, optical and ultraviolet data. An even wider range of
galaxy properties were discussed by other speakers.

{\bf Cluster membership:} Looking beyond the cluster core, {\sc
Gavazzi} noted that HI deficiency is a good indicator of cluster
membership, and that by this criterion a large fraction of the spirals
in the Coma region are cluster (or supercluster) members. {\sc
Bravo-Alfaro} examined the possible HI removal mechanisms and
concluded that ram-pressure stripping is likely to be dominant. On the
other hand, {\sc Boselli} noted that H$_2$ is {\em not} removed,
apparently because it generally lies much deeper down the galaxy
potential wells.

{\bf Luminosity functions:} Several speakers showed new data on the
cluster luminosity function (LF), both in the optical ({\sc Gavazzi},
{\sc Lobo}, {\sc Sekiguchi}, {\sc De Propris}) and the near infrared
({\sc Gavazzi}, {\sc De Propris}). In the optical there now seems to
be agreement that there are three LF regimes: (1)~galaxies brighter
than $L^*$ occupy an exponential tail to high luminosities;
(2)~sub-$L^*$ galaxies follow a power law with slope
$\alpha$$\approx$$-$1.25 ({\sc Katgert}); while (3)~dwarfs cause a
steep upturn at the faint end of the LF, with power law slope
somewhere in the range $-$1.3$>$$\alpha$$>$$-$2. The overall shape
might be represented by either a bright-end Schechter function plus a
faint-end power law or by a bright-end Gaussian plus a faint-end
Schechter function. However {\sc Lobo} noted (as in some previous
work) that the sub-$L^*$ range in Coma shows a significant bump and
dip rather than a simple power-law. {\sc Sekiguchi} examined the LF
variations with radius within Coma and concluded that the faint-end
(`dwarf') LF steepened from $\alpha$$\approx$$-$1.5 in the cluster
core to $\alpha$$\approx$$-$1.9 beyond one Abell radius, a result
consistent with the dwarf `deficit' noted in the core by {\sc Lobo}.

Comparing the optical cluster LF with the field LF in each of these
regimes shows that: (1)~at the bright end the LFs are similar except
for the presence of D/cD galaxies in some clusters and (perhaps) a
slightly brighter $L^*$; (2)~the sub-$L^*$ LFs are also similar,
though clusters show a somewhat steeper slope than the field, at
$\alpha$$\approx$$-$1.25 rather than $\alpha$$\approx$$-$1.0; (3)~the
very faint end of the LFs cannot really be compared since this regime
is effectively unmeasured in the field due to the very small volumes
encompassed by magnitude-selected samples at these luminosities. Large
and/or deep redshift surveys may help here (such as the Sloan or 2dF
wide-angle surveys or the deep survey reported in the poster by {\sc
Adami}), but it will require much work to accurately pin down the LF
fainter than $M^*$+4.

Consideration of the near-infrared (NIR) LF is motivated by the
observation that the H or K band luminosity is the best estimator of
stellar mass ({\sc Gavazzi}, {\sc De Propris}). In fact the H band LF
has a similar overall shape to the optical LF, implying that the faint
end of the optical LF is not entirely due to elevated star-formation
amongst dwarf galaxies---there really are lots of low-mass galaxies,
though they tend to be blue. The H band LFs in the cluster and the
field are similar, at least within the current measurement
uncertainties.

{\bf Colour--magnitude relation:} {\sc Secker} presented a B$-$R {\it
vs} R colour--magnitude (CM) diagram which shows that the standard CM
relation extends all the way down to $M_R$$\approx$$-13$. Like {\sc
Lobo} and {\sc Sekiguchi}, he noted the deficit of faint dE's in the
cluster core.  {\sc Terlevich} noted a small dependence of the
zeropoint of the U$-$V CM relation on cluster radius, with the core
having a zeropoint $\sim$0.1\,mag redder than the periphery. Although
this could possibly be due to different star-formation histories or a
variation in the cluster dust content with radius, it may simply
reflect the morphology--density relation, since {\sc De Propris} found
slightly different CM relations for E's and S0's (with the E's, more
common in the core, having a flatter relation than S0's, which
dominate the CM relation at larger radii).

{\bf Morphology and environment:} {\sc Andreon} raised the interesting
question of whether morphological segregation is primarily based on a
privileged cluster direction or axis rather than density or
clustocentric radius.  He pointed to the highly elongated distribution
of early-type galaxies in the cluster running along the axis through
the two central dominant galaxies, as previously emphasised by {\sc
West}. In contrast the spirals show a far more diffuse and isotropic
distribution. This difference was also noted by {\sc Nichol} in
comparing the distributions of the reddest and bluest galaxies in the
cluster. Is the formation of early-type galaxies more common either in
subclusters or subcluster mergers, so that the early types show a
kinematic echo of the orbits of their parent subclusters?

{\bf Fundamental Plane:} {\sc Mobasher} showed some very striking
results based on his study of the near-infrared fundamental plane
(FP). One of the main motivations for using the NIR FP is that it
should be less subject to variations in star-formation history since
the NIR light is dominated by the old stellar population. For this
reason his K-band FP is puzzling in two respects: firstly, he finds a
slope log$D$/log$\sigma$ of 1.2, which is even further from the virial
theorem slope of 2 than the optical FP (for which the slope is
$\sim$1.4) and disagrees with the K-band slope of 1.7 found by Pahre
\& Djorgovski (1997); secondly, his NIR FP has virtually identical
scatter to the optical FP, when it might be expected to be
less. Setting aside these issues, however, {\sc Mobasher} finds a
quite startling difference in the NIR FPs he obtains for Coma and the
Great Attractor (GA) region, with the GA ellipticals having diameters
which are 10--20\% smaller at fixed velocity dispersion. If correct,
this would significantly reduce the Local Group infall into the GA;
however it would also imply large variations in galaxy properties
correlated on scales of at least tens of Mpc. So we have a choice:
either large-scale coherent flow or large-scale correlated variations
in galaxy formation! Since Pahre \& Djorgovski (1997) do not see such
large variations in the NIR FP in comparing different clusters, it is
clearly essential to confirm these preliminary results.

{\bf Ages and Metallicities:} {\sc Mehlert} has performed spectroscopy
on 35 early-type galaxies in Coma which she has used to investigate
the ages and metallicities of this population, and also correlations
of metallicity with galaxy luminosity and radius. Within the framework
(and to the accuracy) provided by the models of Worthey (1994), she
was able to conclude that the E's are all old ($>$8\,Gyr), while the
S0's have a wide range of effective ages. Intriguingly, the two
central dominant galaxies have higher metallicities than the S0's but
lower effective ages than the E's. This type of analysis could prove
very revealing if the models can be shown to be reliable---for
example, the models need to account for the over-abundance of Mg
compared to Fe, and to explain why more massive E's have a greater
over-abundance than less massive E's. Another, perhaps related,
challenge for models of elliptical galaxy formation is to reproduce
the observed radial metallicity gradients and account for why steeper
gradients are found for more massive galaxies.

{\bf The Butcher-Oemler Effect:} {\sc Caldwell} and {\sc Rose}
presented the results of studies into the origin of the Butcher-Oemler
(BO) effect and the nature of post-starburst (PSB) galaxies in Coma
and other nearby clusters. First, they confirm using diagnostic
spectral lines that PSB galaxies have indeed undergone a burst of
star-formation and are not simply the result of truncating a
previously constant star-formation rate. Secondly, they do not find
evidence that PSB galaxies are the result of pairwise
interactions---they speculated that the trigger may instead be global
tidal effects induced either by infall onto the cluster or by
subcluster mergers. An important question was raised by both {\sc Jones}
and {\sc van Haarlem}: is infall onto the cluster dominated by single
galaxies or by bound groups?

\section{Simulations}

The question that must be asked of the growing hordes simulating every
possible astrophysical process and situation is: How virtual is your
reality? Are the simulations (as we hope) leading to new physical
insights and revealing the outcomes of processes too complicated to
follow analytically? There are two opposite sorts of problems commonly
encountered: either the simulations, for whatever reason, fail to
adequately reflect reality (they are too `virtual'), or they require
so much calibration from observation that they have no option but to
reflect their empirical inputs (they are forcibly constrained to be
`real'). A `good' simulation is as virtual as possible (has few
calibrating empirical inputs) while remaining as close as possible to
reality.

In his review, {\sc Evrard} pointed to the very significant increase
in the sophistication and reliability of the current generation of
cluster simulations, in particular the advance from purely N-body
gravitational simulations of the mass (dark matter halos) to
simulations incorporating not only the gas dynamics but also
preliminary recipes for star formation and other messy galaxy
processes. He suggested that in fact we have reached, or even passed,
the break-even point where the observational input is exceeded by the
predictive output. Although the simulations at this level have yet to
mature, he listed some results that he felt were robust, including:
(i)~galaxies at $r<r_{200}$ are in virial equilibrium
($\rho/\rho_{crit}$=200 at $r_{200}$); (ii)~the velocity anisotropy
$\beta$ is nearly zero at 0.1$r_{200}$ and increases to 0.5 at
$r_{200}$; (iii)~velocity bias is small --
$\sigma_{gal}/\sigma_{DM}\approx0.75$.

{\sc Evrard} presented two other interesting results relating to the
baryon fraction and $\Omega$. First, from a sample of 19 clusters
including Coma, he derives a typical gas mass fraction within
$r_{500}$ of (0.06$\pm$0.03)$h^{-3/2}$, implying either that $\Omega_M
h^{2/3}$=0.07--0.28 (for low to high D/H ratio), or that we don't
understand Big Bang nucleosynthesis. Second, a method which compares
the mass density contrast $\delta_C$ (from $T_X$) and the galaxy
density contrast $\delta_N$ (from galaxy counts) gives
$\Omega_0/b_{cluster}=\delta_C/\delta_N\sim0.3$, similar to the
estimate obtained from the standard optical $M/L$ ratios for clusters.

{\sc van Kampen} presented a summary of the techniques for constrained
random field realizations of specific clusters. These methods are in
their infancy as yet, but hold out the tempting prospect of a direct
simulation attack on the merger histories and perhaps even (if star
formation can be modelled reliably) the galaxy properties of specific
clusters.

Descending in scale from simulations of rich clusters to simulations
of small groups, {\sc Athanassoula} showed the results of some
investigations aimed at understanding the processes and initial
conditions which lead to the formation of dominant brightest cluster
galaxies (BCGs). While no inevitable sequence emerges from what is a
necessarily stochastic process, some interesting heuristic rules seem
to apply: (i)~a BCG nearly always forms in poor cluster simulations
unless there is nothing resembling a suitable central `seed'
concentration from which to grow; (ii)~the mass in BCGs acquired by
cannibalism and mergers is generally larger than (or at least
comparable to) the mass acquired by accretion of material stripped
from other galaxies; (iii)~central concentration is a more important
determinant of BCG formation than the total mass of the common halo;
(iv)~a cD-like extended halo is formed when accretion of stripped
material is (unusually) the dominant process. In order to confront
these results with observations, such simulations need to be combined
with merger history trees for clusters in order to make predictions
for the properties of BCGs as a function of cluster properties such as
mass/richness and degree of substructure.

\section{Hot Gas, Warm Gas, Cool Dust}

The final session of the meeting dealt with X-ray, UV and far-infrared
(FIR) observations of Coma, and demonstrated that a lot can be learnt
from very few photons.

{\sc Jones} summarised the now well-understood main properties of the
X-ray gas, noting the greatest remaining deficiency in the
observational picture is the lack of high-resolution temperature
maps. In X-rays, Coma is a fairly normal massive cluster. Nonetheless,
as in the optical, ``the better the observations, the more
substructure''. The wavelet analysis of the ROSAT X-ray image by
Vikhlinin \etal\ (1997) shows a clear double core surrounded by a more
diffuse component (in which respect it is more than reminiscent of the
wavelet analyses of the galaxy distribution by {\sc Gerbal} and {\sc
Slezak} discussed above). It also shows a distinct tail trailing from
NGC\,4911 towards the cluster core. A preliminary temperature map of
the core is very messy (as expected from the simulations) and hard to
interpret. Interestingly, the NGC\,4839 group has a high X-ray
temperature for its velocity dispersion, such as is also seen in the
merging Centaurus clusters.

Temperature maps of the whole cluster out to about 1\,degree from the
centre were shown by both {\sc Briel} and {\sc Honda}. While these
maps have very low spatial resolution (tens of arcmin) and are clearly
only a first step, nonetheless the two maps did show broad agreement
in their main features. {\sc Briel} noted that the gas is hot between
the cluster core and the NGC\,4839 group, while {\sc Honda} emphasised
a decreasing temperature gradient from the NW to the SE of the
cluster.

An update on mass estimates for Coma was given by {\sc Hughes}, based
on improved X-ray and optical data and modelling. He finds that the
gas (and perhaps also the galaxies) have a more extended distribution
than the dark matter. For $h$=0.5, the mass inside 1,3,5\,Mpc is in the
range 5--7,8--19,10--27$\times$10$^{14}$M$_{\odot}$ (the inner two
estimates are considered reliable, while the outer one is more
uncertain). The baryon fraction, estimated inside 1,3\,Mpc is
13--17\%,18--43\%.

Dropping in temperature from the X-ray to the UV, {\sc Lieu} presented
a very careful case for the existence of a significant amount of
sub-10$^6$\,K gas in a number of clusters. In Coma, the EUVE satellite
detects emission in the 50--190\AA\ band out to 15\,arcmin from the
cluster centre. This represents an excess over the flux predicted from
the 8\,keV X-ray gas. {\sc Lieu} claims that a three-phase model is
needed, with gas at temperatures of 8\,keV, 0.3\,keV and 0.07\,keV. The
latter component has a cooling time of only $\sim$1\,Gyr, so that
10$^{14}$\,M$_{\odot}$ should have cooled out over a Hubble time---where
has this material gone?

Descending even further in temperature, {\sc Stickel} reported the
first direct detection of dust in a cluster---the ISOPHOT instrument
on the ISO satellite detects the signal of dust with a temperature of
$\sim$30\,K in Coma. The mass involved is only 10$^8$\,M$_{\odot}$,
consistent with previous upper limits on the dust content of the
cluster. The gas/dust ratio is low compared to the Galaxy, yet since
the dust would be destroyed by sputtering in $\sim$0.1\,Gyr it is not
primordial, perhaps resulting from galactic winds or stripping of the
ISM.

\section{In Sum}

I have been inspired, intrigued and enthused by so much I have heard
during this conference. I offer my thanks, and the thanks of all
the participants, to Alain Mazure, Florence Durret, Daniel Gerbal and
Fabienne Casoli for organising this most successful meeting, and to
the various institutions and organisations which sponsored it.

\end{document}